\documentclass[12pt]{iopart}
\usepackage{iopams}
\usepackage{amsfonts}
\usepackage{amssymb}
\usepackage{graphicx}
\usepackage{xcolor}
\usepackage{cite}

\newcommand{\bsx}{\boldsymbol{x}}
\newcommand{\bsu}{\boldsymbol{u}}
\newcommand{\bsp}{\hat{\boldsymbol{p}}}
\newcommand{\bsS}{\boldsymbol{S}}

\newcommand{\dd}{\mathrm{d}}
\newcommand{\vs}{v_\mathrm{s}}
\newcommand{\veff}{v_\mathrm{eff}}
\newcommand{\vmax}{v_\mathrm{max}}
\newcommand{\rmax}{r_0}
\newcommand{\umax}{u_0}
\newcommand{\XFP}{X_\mathrm{FP}}
\newcommand{\PFP}{P_\mathrm{FP}}

\begin{document}

\title{
Microswimmers in an axisymmetric vortex flow
}

\author{Jos\'{e}-Agust\'{i}n Arguedas-Leiva$^{1,2}$ and Michael Wilczek$^{1,2}$}
\address{$^{1}$Max Planck Institute for Dynamics and Self-Organization (MPI DS), Am Fa{\ss}berg 17, 37077 G\"{o}ttingen, Germany}
\address{$^{2}$Faculty of Physics, University of G\"{o}ttingen, Friedrich-Hund-Platz 1, 37077 G\"{o}ttingen, Germany}
\eads{\mailto{michael.wilczek@ds.mpg.de}}

\date{\today}

\begin{abstract}
Microswimmers are encountered in a wide variety of biophysical settings. When interacting with flow fields, they show interesting dynamical features such as trapping, clustering, and preferential orientation. One important step towards the understanding of such features is to clarify the interplay of hydrodynamic flows with microswimmer motility and shape. Here, we study the dynamics of ellipsoidal microswimmers in a two-dimensional axisymmetric vortex flow. Despite this simple setting, we find surprisingly rich dynamics, which can be comprehensively characterized in the framework of dynamical systems theory. By classifying the fixed-point structure of the underlying phase space as a function of motility and microswimmer shape, we uncover the topology of the phase space and determine the conditions under which microswimmers are trapped in the vortex. For spherical microswimmers, we identify Hamiltonian dynamics, which are broken for microswimmers of a different shape. We find that prolate ellipsoidal microswimmers tend to align parallel to the velocity field, while oblate microswimmers tend to remain perpendicular to it. Additionally, we find that rotational noise allows microswimmers to escape the vortex with an enhanced escape rate close to the system's saddle point. Our results clarify the role of shape and motility on the occurrence of preferential concentration and clustering and provide a starting point to understand the dynamics in more complex flows.
\end{abstract}

\submitto{\NJP}

\maketitle

Driven by the need to better understand the physical mechanisms of microswimming, e.g.~in the context of phytoplankton in the ocean \cite{guasto2012arfm} or artificial microswimmers in the laboratory \cite{kim2012microbiorobotics}, the investigation of microswimmers in complex flows has gained considerable momentum over the past years. In such flows, inertial effects \cite{schmitt2008jms,wang2012jfm,wang2012pof,li2016pre,felderhof2019ejm}, gyrotactic swimming \cite{
bearon2011jfm,
durham2011prl,
durham2013natcom,
delillo2014prl,
santamaria2014pof,
zhan2014jfm,
pedley2015jfm,
gustavsson2016prl,
borgnino2017pre,
borgnino2018jfm,
richardson2018prf,
cencini2019epje
}, fluid-cell \cite{prairie2012limocean,khurana2012pof} and cell-cell interaction \cite{khurana2013njp,breier2018pnas}, as well as active motility and morphological changes \cite{michalec2017pnas,sengupta2017nat} can give rise to complex spatial microswimmer distributions and enable migration strategies. 

Among the various mechanisms, shape and motility are the key parameters in quantifying microswimmer interaction with hydrodynamic flows \cite{fouxon2015pre,pujara2018jfm,niazi2017jfm,borgnino2019prl}. For instance, motility is a crucial ingredient for the emergence of clustering of neutrally buoyant particles \cite{zhan2014jfm,borgnino2017pre,pujara2018jfm}. Shape, on the other hand, determines the dynamic reaction to hydrodynamic cues. Rod-like agents, for example, tend to align with the direction of local vorticity, while also spinning due to it \cite{pumir2011njp,chevillard2013,ni2014jfm,parsa2012prl,pujara2017jfm}. Furthermore, it has recently been shown that self-propelled rod-shaped particles tend to align with the velocity field \cite{borgnino2019prl}.

Given the complexity of turbulent and spatiotemporally varying flows, several investigations have focused on simple flows, which allow to make contact with dynamical systems theory. For example, the (quasi-)periodic \cite{zoettl2012prl} and chaotic \cite{chacon2013pre} motion of spherical microswimmers in a Poiseuille flow can be explained through the underlying Hamiltonian dynamics. In this scenario, the effect of shape in microswimmer dynamics and its relation to the swimming velocity could be well understood: Elongated microswimmers explicitly break the symplectic structure of the dynamics, but their swimming behavior qualitatively follows the swinging and tumbling motion observed for spherical microswimmers \cite{zoettl2013epje}. As another example, the ubiquity of vortices in natural environments and their dynamical impact on biological agents \cite{sokolov2016natcom} continue to make the study of single vortex structures a good starting point for understanding microswimming in more complex, biologically relevant flows \cite{jumars2009marec}. Isolating the impact of individual vortices on microswimming in a two-dimensional cellular flow has revealed barriers to particle transport as a function of shape and swimming speed \cite{torney2007prl,khurana2011prl}. Shape deformation, as another example, has been found to have a strong impact on the scattering dynamics of individual microswimmers in a single vortex structure \cite{tarama2014pre}. This illustrates how the investigation of simple flow settings sheds light on the interplay between shape and swimming speed on microswimmer dynamics.

Here, we comprehensively characterize microswimmer dynamics in a single vortex structure by relating the observed physical phenomena to properties of the underlying dynamical system. In particular, we consider non-interacting microswimmers in a two-dimensional axisymmetric vortex flow and address their trapping properties as well as the occurrence of clustering (the spatially heterogeneous distribution of particles) and preferential orientation with respect to the velocity field. We idealize microswimmers as advected ellipsoidal particles, which additionally have a swimming direction and a constant self-propulsion speed. Furthermore, vorticity and shear induce particle tumbling, which alters the swimming direction.
Our study discusses the fundamental dynamical systems properties of ellipsoidal microswimmers in a general axisymmetric vortex flow, complementing previous works on simple vortex \cite{sokolov2016natcom} and cellular flows \cite{torney2007prl}.

This simple setting reveals surprising insights: we identify an effective swimming velocity, which takes into account both motility and shape, as a control parameter for this system. We find surprisingly rich dynamics, which can be comprehensively characterized in the framework of dynamical systems theory. Classifying the fixed-point structure of the underlying phase space as a function of the effective swimming velocity allows to distinguish microswimmers that escape the vortex from the ones which remain trapped. Moreover, we find that spherical microswimmers obey Hamiltonian dynamics, whereas other shapes break the symplectic structure of phase space. Hence phase-space contraction, and ultimately preferential concentration, can be set into the context of breaking of Hamiltonian dynamics by departure from a spherical shape. Finally, to quantify the robustness of our results, we investigate the impact of rotational noise. We find that a saddle point, present in the relevant phase space, plays an important role in the escape of microswimmers from the vortex core.

\section{Microswimmers in a vortex flow}
\subsection{Model equations}

We model microswimmers as inertialess particles advected by a velocity field $\bsu$. The particles are additionally capable of self-propulsion with swimming velocity $\vs$ in direction $\bsp$ \cite{pedley1992arfm}. The microswimmer position $\bsx$ obeys the equation of motion
\begin{equation}
\dot{\bsx}=\bsu+\vs\:\bsp,
\label{eq:eom_x}
\end{equation}
where the flow field is evaluated at the Lagrangian position of the microswimmer $\bsu(t,\bsx(t))$. A simple but effective way to introduce shape is to consider ellipsoidal microswimmers \cite{kaya2009prl}. Particle orientation can be described by the particle's symmetry axis $\bsp$. Additionally, in many relevant settings microswimmers are smaller than the smallest hydrodynamic scales (such as the Kolmogorov length scale in turbulence). In this limit, the spinning and tumbling of the particle orientation can be described by Jeffery's equation \cite{jeffery1992prsla}, which takes the form
\begin{equation}
\dot{\bsp}=\frac{1}{2}\bomega\times\bsp+\alpha(\bsS\bsp-\bsp^\mathrm{T} \bsS\bsp\:\bsp).
\label{eq:eom_p}
\end{equation}
Here, $S_{ij}=(\partial_iu_j+\partial_ju_i)/2$ is the strain tensor and $\bomega=\nabla\times\bsu$ the vorticity. The ratio between the ellipsoid's major and minor axes $\lambda$ defines the shape parameter as $\alpha=(\lambda^2-1)/(\lambda^2+1)$. The parameter $\alpha$ interpolates shapes between an oblate ellipsoid ($-1<\alpha<0$), a sphere ($\alpha=0$), or a prolate ellipsoid ($0<\alpha<1$).
\subsection{Two-dimensional vortex flow}

\begin{figure}
\centering
\includegraphics[width=1.0\linewidth]{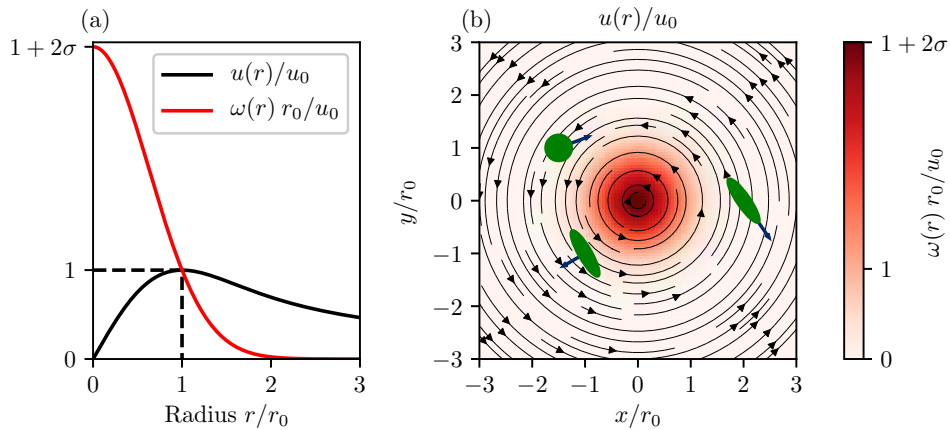}
\caption{To illustrate our results, we choose a Gaussian vortex, the stationary Lamb-Oseen vortex \eref{eq:lamb_oseen}. (a): Velocity and vorticity profiles as a function of radius. (b): Streamline plot of the velocity field with vorticity shown as color-coded background. In our system ellipsoidal microswimmers can self-propel and are also advected by the vortex flow.
}
\label{fig:figure1a_1b}
\end{figure}

In two spatial dimensions, \eref{eq:eom_x} and \eref{eq:eom_p} have three degrees of freedom: two coordinates for position and one swimming angle
\begin{equation}
\bsx=\left(\begin{array}{c} x \\ y\end{array}\right), \quad \bsp=\left(\begin{array}{c}\cos\theta \\ \sin\theta\end{array}\right).
\label{eq:parametrization_xp}
\end{equation}
It is also worth noting that in two dimensions ellipsoids can only tumble. In the following, we restrict ourselves to an axisymmetric vortex flow of the form $\bsu(t,\bsx)=u(r)\: \hat{\textbf{e}}_\phi$, where $r=\sqrt{x^2+y^2}$ is the radial coordinate and $\hat{\textbf{e}}_\phi$ a unit vector along the angular coordinate. By inserting \eref{eq:parametrization_xp} into \eref{eq:eom_x} and \eref{eq:eom_p}, we obtain the equations of motion
\begin{eqnarray}
\dot{x}=&v_s\cos\theta-\frac{u(r)}{r}y, \label{eq:eom_xytheta1} \\
\dot{y}=&v_s\sin\theta+\frac{u(r)}{r}x, \label{eq:eom_xytheta2} \\
\dot{\theta}=&\frac{1}{2r}\frac{\dd}{\dd r}\left[ r\: u(r) \right]+\alpha\frac{1}{2r}\frac{\dd }{\dd r} \left[\frac{u(r)}{r}\right]  \label{eq:eom_xytheta3}
\\ &  \times\left[(x^2-y^2)\: \cos (2\theta)+ (2xy)\:\sin (2\theta)\right] .\nonumber
\end{eqnarray}
Inspection of these equations in polar coordinates reveals rotational invariance. Without loss of generality, the dynamics of the system can therefore be reduced by one degree of freedom by transforming into a co-rotating frame. This is achieved by introducing new coordinates
\begin{equation}
\left(\begin{array}{c} X \\ P\end{array}\right)=\left(\begin{array}{rr} \cos\theta & \sin\theta \\ -\sin\theta & \cos \theta\end{array}\right)\left(\begin{array}{c} x \\ y\end{array}\right).
\label{eq:parametrization_XP}
\end{equation}
In the co-rotating coordinate system $\lbrace X,P\rbrace$, the microswimmers are rotated so that they swim along the positive $X$ axis. That is, in this new frame the microswimmer position and orientation differ only by a rotation from the laboratory frame. Hence, the radial coordinate is identical in both coordinate systems,
\begin{equation}
r=\sqrt{x^2+y^2}=\sqrt{X^2+P^2}.
\end{equation}
Introducing an angle variable defined as $\tan\psi:=P/X$ helps in the description of the dynamics. The equations of motion for $X$ and $P$ can be obtained from \eref{eq:eom_xytheta1}-\eref{eq:parametrization_XP} as
\begin{eqnarray}
\dot{X}=&\vs+P\: f(r)\:\left[1+\alpha\: h(\psi)\right], \label{eq:eom_XP1} \\
\dot{P}=&-X\:f(r)\:\left[1+\alpha\: h(\psi)\right], \label{eq:eom_XP2}
\end{eqnarray}
where
\begin{equation}
f(r):=\frac{r}{2}\frac{\dd}{\dd r}\left[\frac{u(r)}{r}\right] \quad \textmd{and} \quad h(\psi):=\cos(2\psi).
\end{equation}
The function $f(r)$ contains the dependence on the velocity field $u(r)$. Vorticity and strain alter microswimmer orientation. While vorticity affects all microswimmers independent of their shape, the effect of strain is shape-dependent. The function $h(\psi)$ is a geometric term stemming from the relative orientation of the ellipsoid with respect to the strain tensor.\newline
The swimming orientation angle $\theta$ is a slave variable to $X$ and $P$ and evolves according to the equation
\begin{equation}
\dot{\theta}=\frac{u(r)}{r}+f(r)\:\left[1+\alpha\: h(\psi)\right].
\label{eq:eom_theta}
\end{equation}
Hence, this co-rotating representation reduces the three-dimensional dynamics of the microswimmers in the laboratory frame $\lbrace x,y,\theta\rbrace$ to an effective two-dimensional dynamics in the co-rotating frame $\lbrace X,P\rbrace$. In general, a particle's squared tumbling rate is defined as $\dot{\bsp}^2$. In our parametrization \eref{eq:parametrization_xp}, the squared tumbling rate simplifies to $\dot{\theta}^2$, and  therefore can be directly determined from the swimming-angle dynamics \eref{eq:eom_theta}. For spherical microswimmers, the tumbling rate only depends on the radial coordinate, while \eref{eq:eom_theta} shows that strain additionally induces orientation-dependent tumbling for ellipsoidal microswimmers.
\newline
As a concrete example, we consider a stationary Lamb-Oseen vortex, a prototypical vortex structure which is representative for a large class of hydrodynamic vortices \cite{wu2007vorticity}. The essential feature of this flow field is that the velocity profile interpolates between linear growth near the core (corresponding to a solid body rotation) and a Gaussian decay far from the core, leading to differential rotation. Its azimuthal velocity is given by
\begin{equation}
u(r) = \umax\frac{\rmax}{r}\left(1+\frac{1}{2\sigma}\right)\left(1-\exp\left[\frac{-\sigma\: r^2}{{\rmax}^2}\right]\right)
\label{eq:lamb_oseen}
\end{equation}
where $\umax$ and $\rmax$ are the maximum azimuthal velocity and its corresponding radial coordinate, respectively. The vortex is stationary, and hence its width $\rmax$ is kept fixed. The constant $\sigma$ is the nontrivial solution to the equation $\exp(\sigma)=(1+2\sigma)$ \cite{devenport1996jfm}, which is obtained by fixing $\umax$ to $\rmax$. The velocity and vorticity profiles are illustrated in \fref{fig:figure1a_1b}.
\begin{figure}
\centering
\includegraphics[width=1.0\linewidth]{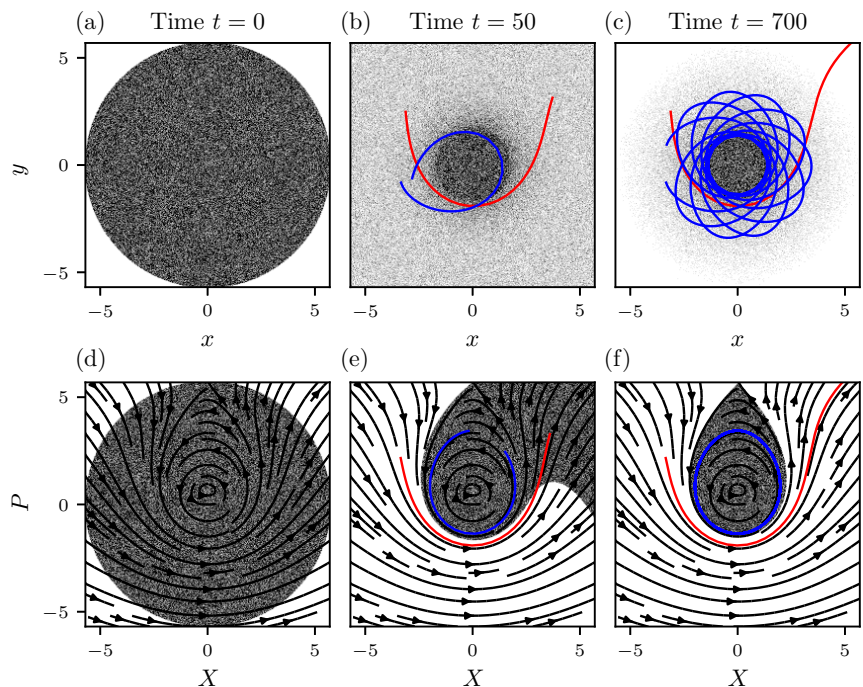}
\caption{(a)-(c): A spherical ($\alpha=0$) microswimmer ensemble (gray density plot) homogeneously initialized in a disk in the laboratory frame $\lbrace x,y,\theta\rbrace$ quickly separates into bound and unbound microswimmers. Example trajectories of bound and unbound microswimmers are shown in blue and red, respectively. Trapped microswimmers remain near the vortex core and follow quasi-periodic orbits. A non-trivial microswimmer distribution develops. (d)-(f): An analysis in the co-rotating frame $\lbrace X,P\rbrace$ reveals the phase-space structure of the dynamics (see figure \fref{fig:figure3a_3c}).}
\label{fig:figure2a_2f}
\end{figure}\newline
Using $\rmax$ and $\rmax/\umax$ as length and time scales, respectively, \eref{eq:eom_xytheta1}-\eref{eq:eom_xytheta3} and \eref{eq:eom_XP1}-\eref{eq:lamb_oseen} can be non-dimensionalized: $x\rightarrow \rmax\:\tilde{x}$, $y\rightarrow \rmax\:\tilde{y}$, $X\rightarrow \rmax\:\tilde{X}$, $P\rightarrow \rmax\:\tilde{P}$, $t\rightarrow \rmax/\umax\:\tilde{t}$, $u\rightarrow \umax\:\tilde{u}$. As a result, we have a non-dimensional swimming velocity $\tilde{v}_s=v_s/\umax$. In the following, we drop the tildes and work with the non-dimensionalized swimming speed. 
\newline
For the numerical results, we integrate the ordinary differential equations \eref{eq:eom_xytheta1}-\eref{eq:eom_xytheta3} using a fourth-order Runge-Kutta method with a time step $\dd t=0.05$. For each of the simulations shown in \fref{fig:figure2a_2f} and \fref{fig:figure5a_5f}, we initialized $5\times 10^5$ microswimmers. For the radial distribution functions in \fref{fig:figure6a_6b}, we initialized $8\times 10^6$ microswimmers.\newline
Typical dynamics of microswimmers following \eref{eq:eom_xytheta1}-\eref{eq:eom_xytheta3} and the corresponding representation in the co-rotating frame $\lbrace X,P\rbrace$ are shown in \fref{fig:figure2a_2f}. Here, the quasi-periodic trajectories observed in the laboratory frame can be explained by the coupling of the typical angular velocity of the microswimmers in the co-rotating frame with their rotation frequency given by \eref{eq:eom_theta}.\newline
In the next section, we explore the features of \eref{eq:eom_XP1} and \eref{eq:eom_XP2} to uncover the fixed-point structure of the underlying dynamics and precisely characterize the observed phenomena of the microswimmers.

\section{Fixed-point analysis}

\begin{figure}
\centering
\includegraphics[width=1.0\linewidth]{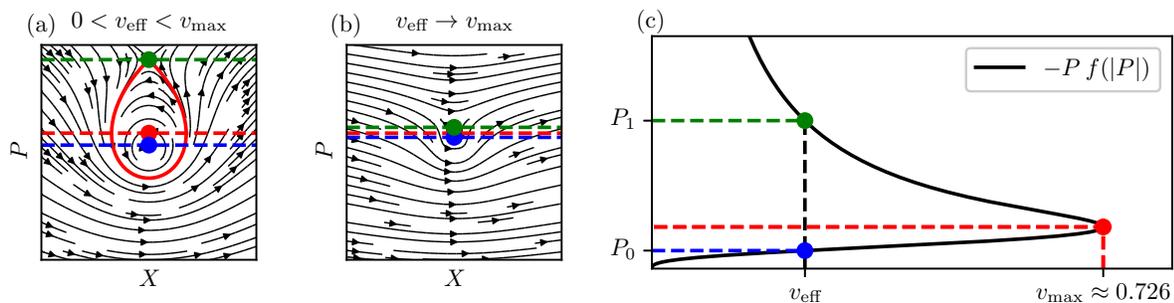}
\caption{
(a) Solving \eref{eq:fixed_points} and classifying the fixed points for microswimmers in a Lamb-Oseen vortex yields a saddle(green)-center(blue) fixed-point pair. The red line indicates the homoclinic orbit. The topology of phase space can be described as a function of an effective swimming velocity $\veff$ \eref{eq:v_eff}. (b) By increasing $\veff$, the fixed points converge and undergo a saddle-node bifurcation at $\vmax$ (red dot). For $\veff>\vmax$, no fixed-point pair exists, and hence no closed trajectories are found.
(c) The bifurcation diagram showing the fixed-point coordinates \eref{eq:fixed_points} reveals the role of $\veff$ as the control parameter for this system.}
\label{fig:figure3a_3c}
\end{figure}

Visual inspection of the microswimmer dynamics in the co-rotating frame in \fref{fig:figure2a_2f} reveals an intricate behavior. These observations can be made precise by a fixed-point analysis. Because the dynamical system \eref{eq:eom_XP1}-\eref{eq:eom_XP2} is two-dimensional, the Poincar\'{e}-Bendixson theorem \cite{strogatz2014nonlinear} applies, and the topology of the dynamics is completely determined by the fixed points. Already in \fref{fig:figure2a_2f} (d)-(f) we can visually identify two fixed points in phase space. These fixed points are marked as blue and green dots in \fref{fig:figure3a_3c} (a) and (b), and are studied in the following.\newline
Calculating the fixed points $\lbrace \XFP,\PFP\rbrace$ of \eref{eq:eom_XP1} and \eref{eq:eom_XP2} trivially leads to $\XFP=0$ for any axisymmetric flow profile. As a consequence, $r=\vert \PFP\vert$ and $\psi=\textmd{sign}(\PFP)\:\pi/2$. For the subsequent analysis, we use the specific example of the Lamb-Oseen vortex \eref{eq:lamb_oseen}, but the theoretical results are valid in general. For the Lamb-Oseen vortex \eref{eq:lamb_oseen}, we find that the inequality $f(r)\leq 0$ is valid for all radii. This means that fixed points exist only for $\PFP>0$ and $\psi=\pi/2$. The fixed points of \eref{eq:eom_XP1} and \eref{eq:eom_XP2} are then given by the solutions to
\begin{equation}
\begin{array}{rl}
0=&\XFP, \\
\frac{\vs}{1-\alpha}=&-\PFP\: f(\vert \PFP\vert).
\end{array}
\label{eq:fixed_points}
\end{equation}
The type of fixed points of a two-dimensional system is determined by the determinant and the trace of its Jacobian matrix $J$ \cite{strogatz2014nonlinear}. At the fixed-point coordinates \eref{eq:fixed_points} we obtain
\begin{equation}
\begin{array}{rl}
\Tr(J)=&0, \\
\det(J)=&\left(1-\alpha\right)^2 f(r)\:\:\frac{\mathrm{d}\left[r\: f(r)\right]}{\mathrm{d}r}\Big\vert_{r=\vert \PFP\vert }.
\end{array}
\label{eq:jacobian}
\end{equation}
These quantities reveal the nature of the fixed points and their dependence on microswimmer shape. Because $\det(J)\in\mathbb{R}$ and $\Tr(J)=0$, this type of system can only have either center or saddle points. As a consequence, trajectories remaining bound to the vortex core correspond to areas in phase space enclosed by the homoclinic or heteroclinic orbits of the saddle points. All other trajectories in phase space lead to and come from infinity. Moreover, \eref{eq:fixed_points} motivates the definition of an effective swimming velocity as
\begin{equation}
\veff:=\frac{\vs}{1-\alpha}.
\label{eq:v_eff}
\end{equation}
As the trace and determinant of $J$ are independent of the swimming speed $\vs$, and the term $(1-\alpha)^2$ in \eref{eq:jacobian} is non-negative, microswimmers with the same effective swimming velocity have identical types of fixed points located at the same coordinates. This implies that the effective swimming velocity can be used to classify microswimmers according to their shape and swimming speed. Therefore, $\veff$ plays the role of a control parameter for the topology of the phase space. To obtain the bifurcation diagram of this system, the equation
\begin{equation}
\veff=-\PFP\: f(\vert \PFP\vert)
\label{eq:bif_equation}
\end{equation}
\begin{figure}
\centering
\includegraphics[width=1.0\linewidth]{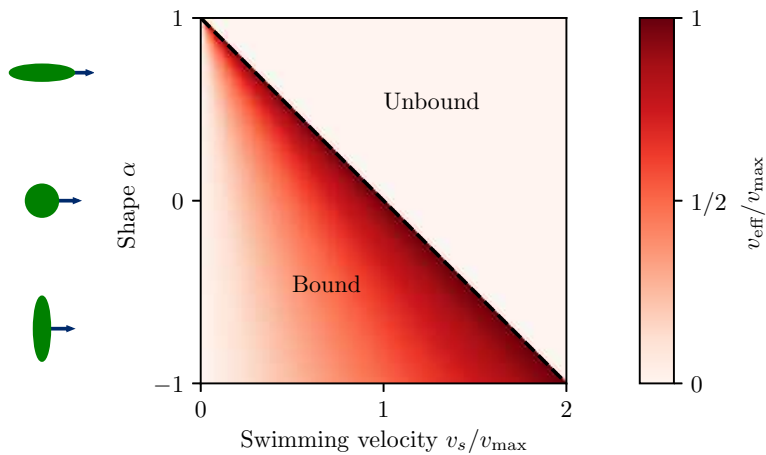}
\caption{
The condition $\veff<\vmax$ defines a region in $\lbrace v_s, \alpha \rbrace$ parameter space for which bound microswimmer trajectories exist. The effective velocity $\veff$ is color-coded and takes the value $\vmax$ at the transition. The transition from bound to unbound microswimmers defines a straight line in $\lbrace \vs,\alpha \rbrace$ parameters space. The transition for prolate microswimmers ($\alpha>0$) occurs at lower $\vs$ than for oblate microswimmers ($\alpha<0$), and hence prolates can more easily escape the vortex core. Furthermore, in the limit $\alpha\rightarrow 1$ the region of bound microswimmers vanishes, and all microswimmers in this shape limit can escape the vortex core.
}\label{fig:figure4}
\end{figure}
can be solved graphically. The solution to this equation for the Lamb-Oseen vortex \eref{eq:lamb_oseen} is shown in \fref{fig:figure3a_3c} (c). In the case of the Lamb-Oseen vortex (\fref{fig:figure3a_3c}) the condition $0<\veff<\vmax$ ensures the existence of a solution pair to \eref{eq:fixed_points}, which corresponds to a center-saddle pair. As long as this solution exists, so does the homoclinic orbit, enclosing a region of trapped microswimmers. \Fref{fig:figure3a_3c} (a) shows the fixed-point pair and the homoclinic orbit. By choosing a larger value for $\veff$, as in \fref{fig:figure3a_3c} (b), the fixed points merge and undergo a saddle-node bifurcation at $\vmax$.\newline
The condition $\veff<\vmax$ defines a region in the $\lbrace v_s,\alpha\rbrace$ parameter space for which microswimmers are trapped, as shown in \fref{fig:figure4}. The transition from bound to unbound microswimmers occurs at $\veff=\vmax$, which defines a straight line in $\lbrace v_s,\alpha\rbrace$ parameter space. For the Lamb-Oseen vortex, we numerically obtain $\vmax\approx 0.726$ as the maximum of \eref{eq:bif_equation}. Hence, by taking into account the role of shape, microswimmers with much lower swimming velocity $\vs$ than the maximum fluid velocity $\umax$ can escape the vortex. Additionally, prolate microswimmers can more easily escape the vortex core than oblate microswimmers. That is, the transition from bound to unbound microswimmers takes place at lower swimming velocity for prolate microswimmers than for oblate microswimmers. Furthermore, for constant $\vs$, thin elongated microswimmers have a divergent $\veff$ in the limit $\alpha\rightarrow 1$, and hence always escape the vortex in this limit.

\section{Hamiltonian dynamics and phase-space contraction}

\begin{figure}
\centering
\includegraphics[width=1.0\linewidth]{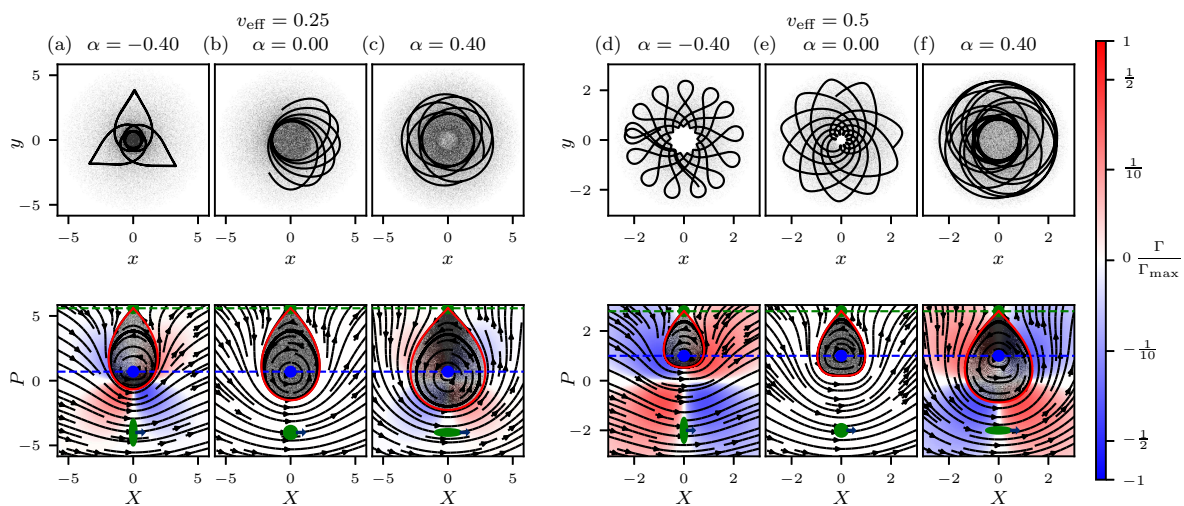}
\caption{
Bound microswimmers develop non-trivial density distributions, shown in the gray scale. We show density distributions of different shapes and effective swimming velocities, namely $\veff=0.25$ in (a)-(c) and $\veff=0.5$ in (d)-(f). The black lines show sample trajectories. For spherical shapes (b,e), we clearly observe a homogeneous distribution in $\lbrace X,P\rbrace$ space, a consequence of the phase-space-conserving Hamiltonian dynamics. A homogeneously filled homoclinic orbit (red line) at the initial time will remain so throughout time evolution. Other shapes break the Hamiltonian structure, and phase-space contraction $\Gamma$ \eref{eq:phase_space_contraction} leads to preferential alignment of microswimmers with respect to the velocity field, as illustrated by the denser regions in $\lbrace X,P\rbrace$ space. Oblate microswimmers (a,d) tend to concentrate along the $X$ axis. Prolate microswimmers (c,f) tend to concentrate along the $P$ axis. This corresponds to preferential swimming perpendicular or parallel to the flow, respectively.
}\label{fig:figure5a_5f}
\end{figure}
Next, we explore the effect of shape on the microswimmer dynamics and its relation to phase-space contraction. Initializing microswimmers homogeneously inside the homoclinic orbit leads to shape-dependent stationary distributions, as shown in \fref{fig:figure5a_5f}. For equal effective swimming velocity $\veff$, changing shape leads to a variety of density distributions.\newline
To elucidate this, we begin by considering spherical microswimmers ($\alpha=0$). It is well known that, in general, equations for spherical microswimmers following \eref{eq:eom_x} and \eref{eq:eom_p} conserve phase-space volume \cite{delillo2014prl,khurana2012pof,borgnino2017pre}
\begin{equation}
\nabla_{\bsx} \cdot \dot{\bsx}+\nabla_{\bsp} \cdot \dot{\bsp}=0.
\end{equation}
An even stronger statement can be made in our case. By using the $\lbrace X,P\rbrace$ coordinates of the co-rotating frame, we can reveal the Hamiltonian structure of the equations of motion, i.e.~
\begin{eqnarray}
\dot{X} &=  \frac{\partial H(X,P)}{\partial P}, \label{eq:ham_dyn_X} \\
\dot{P} &= -\frac{\partial H(X,P)}{\partial X}, \label{eq:ham_dyn_P}
\end{eqnarray}
where the Hamilton function is given by
\begin{equation}
H(X,P)=\vs P+\int^{r} s\:f(s) \: \dd s
\label{eq:hamiltonian}
\end{equation}
for arbitrary axisymmetric velocity fields. As a consequence of the Hamiltonian dynamics, phase-space volume is conserved and does not contract or expand. Therefore, starting from a homogeneous distribution, spherical microswimmers maintain a homogeneous distribution as time evolves.\newline
While spherical microswimmers obey Hamiltonian dynamics, other microswimmer shapes break the Hamiltonian structure. This can be seen by recasting \eref{eq:eom_XP1} and \eref{eq:eom_XP2} as
\begin{eqnarray}
\dot{X}&=\frac{\partial H(X,P)}{\partial P}+\alpha\: P \: f(r)\: h(\psi), \label{eq:ham_dyn_X_alpha}\\
\dot{P}&=-\frac{\partial H(X,P)}{\partial X}-\alpha\: X \: f(r)\: h(\psi).
\label{eq:ham_dyn_P_alpha}
\end{eqnarray}
As a result, phase space can contract or expand for non-spherical microswimmers. We show the stationary distributions of bound microswimmers in both the laboratory frame $\lbrace x,y,\theta\rbrace$ and the co-rotating frame $\lbrace X,P\rbrace$ in \fref{fig:figure5a_5f}. The gray color scale corresponds to microswimmer density. In the co-rotating frame spherical microswimmers remain homogeneously distributed in time. However, non-spherical microswimmers show denser regions inside the homoclinic orbit. Moreover, prolate ellipsoids are more densely concentrated along the $P$ axis. This can be explained by analyzing the phase-space contraction induced by the microswimmer dynamics. The phase-space contraction rate is given by
\begin{equation}
\Gamma:=\partial_X\dot{X}+\partial_P\dot{P}= 2\: \alpha \: \sin(2\psi)\: f(r).
\label{eq:phase_space_contraction}
\end{equation}
Non-spherical microswimmers have an orientation-dependent tumbling rate \eref{eq:eom_theta}, which induces phase-space contraction \eref{eq:phase_space_contraction}. As a consequence, non-spherical microswimmers accumulate or deplete in regions of phase space and will depart from an initially homogeneous distribution. Note that phase-space volume can locally contract and expand. However, integrating over the whole phase space reveals that the total phase-space volume is conserved. The fact that for the Lamb-Oseen vortex \eref{eq:lamb_oseen} the inequality $f(r)\leq 0$ holds together with \eref{eq:phase_space_contraction} implies a constant sign of $\Gamma$ inside each quadrant in $\lbrace X,P\rbrace$ space. As trapped microswimmers traverse the different quadrants, they periodically alternate between expanding and contracting quadrants. Microswimmers exiting an expansion quadrant will show a minimum in density, whilst those exiting a contraction quadrant will show maximum density. Hence, by considering the sign of $\alpha$ and $f(r)$, we conclude that denser regions are formed along the $X$ axis for oblate ellipsoids and along the $P$ axis for prolate ellipsoids.\newline
In \fref{fig:figure5a_5f} the different density regions for oblate and prolate microswimmers can be identified, as well as the homogeneous distribution for spheres. Recall that in the co-rotating coordinate frame $\lbrace X,P\rbrace$, the microswimmers are rotated so that they always point in the positive $X$ direction. The velocity field, on the other hand, is rotationally invariant. Therefore contraction along the different axes reveals that oblate ellipsoids ($\alpha<0$) swim predominantly perpendicular to the velocity field (denser regions along the $X$ axis) while prolate ellipsoids ($\alpha>0$) mostly remain parallel to it (denser regions along the $P$ axis). That means that phase space contracts in such a way that, starting from random initial conditions, trapped microswimmers show shape-dependent preferential orientation (anti-)parallel or perpendicular to the flow. Similar effects have been observed in chaotic, mildly turbulent flows \cite{borgnino2019prl}.\newline
\begin{figure}
\centering
\includegraphics[width=1.00\linewidth]{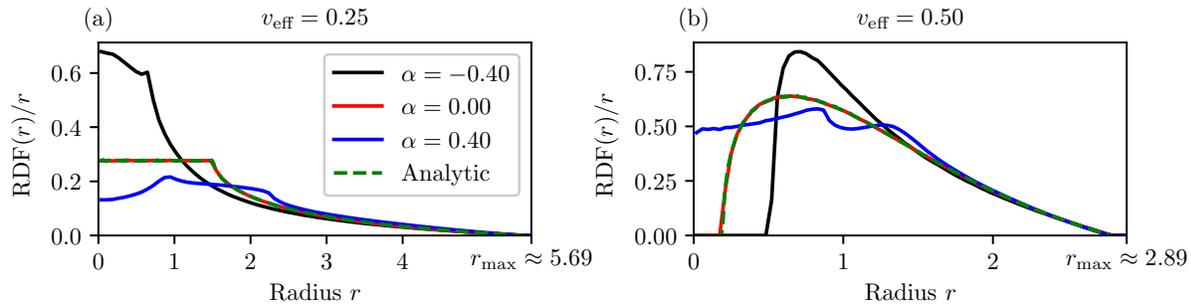}
\caption{The clustering of bound microswimmer ensembles from \fref{fig:figure5a_5f} is characterized by their radial distribution functions (RDFs). Microswimmers with different shape but equal effective swimming velocity have equal maximum swimming radius $r_\mathrm{max}$, given by the saddle point ($P_1$ in \fref{fig:figure3a_3c}). Nevertheless, different shaped microswimmers exhibit different phase-space contraction rates, which leads to different RDFs. (a): For $\veff=0.25$, spherical microswimmers ($\alpha=0$) homogeneously fill an area around the origin. Because the radial distribution function is obtained by an angular integration over $\lbrace X,P \rbrace$ space, this leads to a linearly increasing RDF (i.e.~constant RDF$(r)/r$ ) for circular integration areas fully enclosed by the homoclinic orbit, i.e.~for radii up to $r\approx 1.6$. Beyond this, the RDF starts to decay. Other shapes develop non-trivial RDFs. (b): At higher $\veff$ the vortex core may remain devoid of microswimmers. The analytic line corresponds to homogeneously filling the homoclinic orbit in the spherical case ($\alpha=0$) and integrating out the angle variable. This matches the RDF obtained from the simulations.
}\label{fig:figure6a_6b}
\end{figure}
Interestingly, the dynamical features of the microswimmers in the co-rotating frame lead to clustering in the laboratory frame. To characterize the spatial distribution of microswimmers, we consider the radial distribution function (RDF) (see figure 6). As the laboratory frame and the co-rotating frame $\lbrace X,P\rbrace$ differ only by a rotation, the radial distribution of microswimmer ensembles is identical in both cases. That means that integrating the distribution function in the co-rotating frame over the angle variable exactly corresponds to the RDF obtained in the laboratory frame for any ensemble configuration. In the case of trapped spherical microswimmers ($\alpha=0$), starting from homogeneous initial conditions that fill out the homoclinic orbit (as in \fref{fig:figure2a_2f}), the RDF is unaltered as time evolves. In this case, integrating a constant density inside the homoclinic orbit over the angle variable yields the RDF. Here, it is not necessary to integrate the equations of motion for an ensemble of microswimmers to obtain the stationary RDF; it can be obtained from the shape of the homoclinic orbit alone. For other microswimmer shapes this approach is not feasible, as phase-space contraction sets in under time evolution and a non-trivial stationary distribution develops.\newline
Microswimmers with equal $\veff$ have the same type of fixed points at the same coordinates. Nevertheless, by changing shape, we observe a variety of quasi-periodic orbits and density distributions. These differences are rooted in the shape of the homoclinic orbit as well as the phase-space contraction rate $\Gamma$, which is shown in the background of \fref{fig:figure5a_5f}, normalized by the maximum contraction rate $\Gamma_\mathrm{max}:=2\; \textrm{max}(\vert f(r)\vert )$. Therefore, the RDF of bound microswimmers is a function of both phase-space contraction and the shape of the homoclinic orbit. Both of these differ for microswimmers of different shape and swimming speed, even if their effective swimming speed is identical (see \fref{fig:figure6a_6b}). However, the maximum swimming radius $r_\mathrm{max}$ of trapped microswimmers, beyond which the RDF is zero, is a common property of microswimmers with identical effective swimming speeds. This can be explained by the fact that the saddle point is the point on the homoclinic orbit with the largest radius. Hence the saddle point determines the maximum swimming radius of trapped microswimmers, which is a constant for microswimmers of equal effective swimming speed.

\section{Impact of rotational noise}
\begin{figure}
\centering
\includegraphics[width=\linewidth]{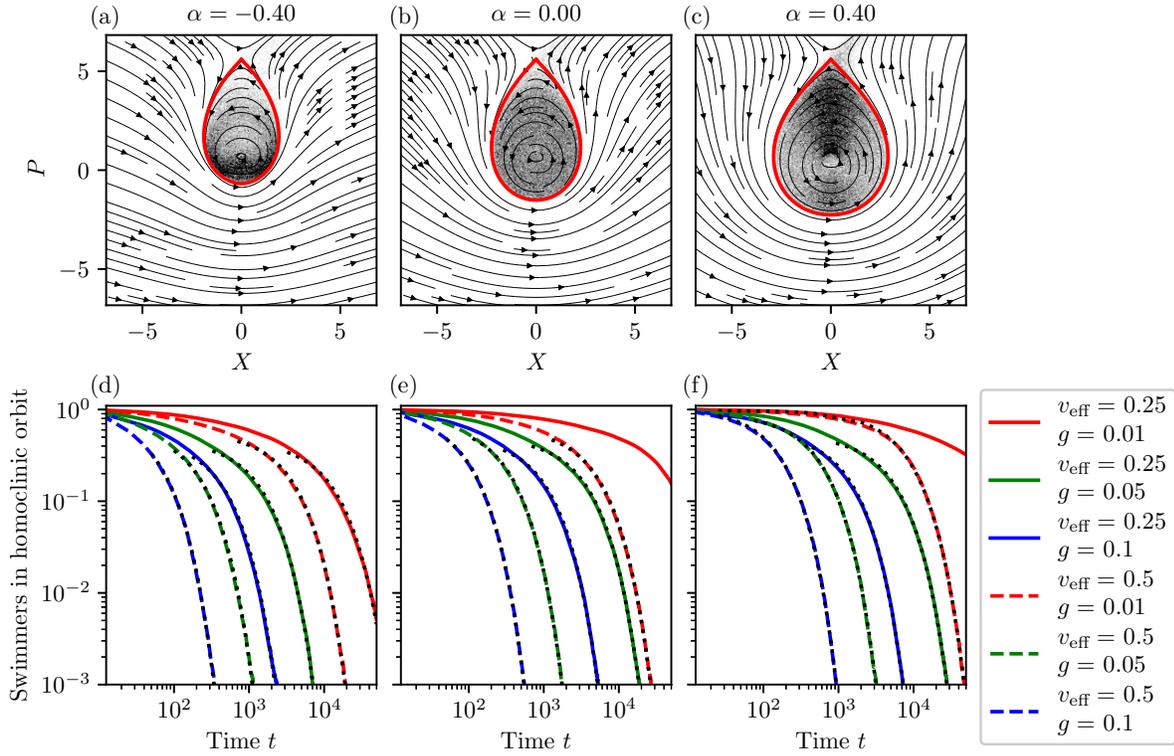}
\caption{(a)-(c) Under the influence of rotational noise microswimmers can escape the vortex core. The saddle point plays an important role here as the maximum expansion direction leads to microswimmers quickly leaving the vortex core. Snapshots taken at time $t=375$ for microswimmers with $\veff=0.25$ and $g=0.01$. (d)-(f) Number of microswimmers inside the homoclinic orbit as a function of time, normalized by the number of microswimmers in the homoclinic orbit at the initial time. For the same $\veff$, prolate microswimmers are slower in escaping the vortex as their $\vs$ is lower than for microswimmers of other shapes. For low values of the effective swimming velocity $\veff$ or rotational noise $g$, the number of microswimmers inside the homoclinic orbit decreases slowly. This is due to the microswimmers only drifting and diffusing slowly towards the saddle point. For higher values of both $\veff$ and $g$, this process happens much more quickly. After an initial transition, the decay can be approximately described as exponential (black dots).
}\label{fig:figure7a_7f}
\end{figure}
So far, we have considered only deterministic microswimmers. In realistic biophysical settings, fluctuations play an important role. We explore this by considering the impact of rotational noise on microswimmers. Let $W$ be a Wiener process, whose increment $\mathrm{d}W$ has zero mean and variance $\dd t$. We then introduce a stochastic term into \eref{eq:eom_p} as
\begin{equation}
\dd\bsp=\dot{\bsp} \, \dd t+\hat{\textbf{e}}_z\times\bsp\: g\: \dd W,
\label{eq:rotational_noise}
\end{equation}
where the deterministic part $\dot{\bsp}$ is given by \eref{eq:eom_p}. Here we are using the non-dimensionalized quantities, such that a P\'{e}clet number can be defined as $\mathrm{Pe}:=1/g^2$. This equation is understood in the Stratonovich sense so that $\bsp$ remains normalized. With the parametrization \eref{eq:parametrization_xp} the equations of motion for microswimmers with rotational noise correspond to additive noise on the swimming angle
\begin{eqnarray}
\dd x &= \dot{x}\:\dd t, \label{eq:noise_x}\\
\dd y &= \dot{y}\:\dd t,  \label{eq:noise_y}\\
\dd\theta &= \dot{\theta}\:\dd t+g\: \dd W, \label{eq:noise_theta}
\end{eqnarray}
where again the deterministic parts $\dot{x}$, $\dot{y}$, and $\dot{\theta}$ are given by \eref{eq:eom_xytheta1}-\eref{eq:eom_xytheta3}. A direct comparison with the deterministic equations reveals that rotational fluctuations enhance the particle's tumbling rate.\newline
For the numerical results presented in this section, we solve \eref{eq:noise_x}-\eref{eq:noise_theta} using the Euler-Maruyama method with a time step $\dd t=0.0005$ and a total of $5\times 10^5$ microswimmers. Switching to the co-rotating $\lbrace X, P \rbrace$ frame, we obtain the stochastic equations (in the Stratonovich sense) as
\begin{eqnarray}
\dd X=\dot{X}\:\dd t+g\:P\:\dd W, \\
\dd P=\dot{P}\:\dd t-g\:X\:\dd W,
\end{eqnarray}
with $\dot{X}$ and $\dot{P}$ given by \eref{eq:eom_XP1} and \eref{eq:eom_XP2}. The effect of noise in the co-rotating frame corresponds to rotational diffusion. This can be seen from the Fokker-Planck equation for the density distribution function $\rho(X,P;t)$
\begin{equation}
\partial_t\rho=-\veff(1-\alpha)\partial_X\rho+\mu\rho-\Gamma\rho+\frac{g^2}{2}\mathrm{D}\rho
\label{eq:Fokker_Planck_equation}
\end{equation}
where $\Gamma$ is the phase-space contraction rate \eref{eq:phase_space_contraction} and $\mu$ is a rotational drift operator
\begin{equation}
\mu:=f(r)[1+\alpha\:h(\psi)](X\partial_P-P\partial_X)= f(r)[1+\alpha\:h(\psi)]\partial_\psi.
\end{equation}
Additionally, $\mathrm{D}$ is a diffusion operator given by
\begin{equation}
\mathrm{D}:=(X\partial_P-P\partial_X)(X\partial_P-P\partial_X)=\partial_\psi\partial_\psi.
\label{eq:diffusion_term}
\end{equation}
Setting $g=0$ in \eref{eq:Fokker_Planck_equation} yields a Liouville equation for the deterministic part of the dynamics \eref{eq:eom_XP1}-\eref{eq:eom_XP2}. For $\vmax<\veff$, the drift term due to swimming dominates, and microswimmers always escape the vortex core. Without swimming ($\veff=0$) only rotational drift is present, i.e.~the microswimmers behave as passive tracers. In the regime $0<\veff<\vmax$ the presence of the fixed-point pair and the homoclinic orbit leads to a stationary solution to the Liouville equation, corresponding to stationary distributions as shown in \fref{fig:figure5a_5f}.\newline
Enhanced tumbling rates due to the addition of rotational noise on the swimming direction lead to rotational diffusion \eref{eq:diffusion_term} in the co-rotating frame. This induces microswimmer transfer across the homoclinic orbit. Therefore, starting from a stationary distribution of microswimmers in the homoclinic orbit, all microswimmers will eventually escape the vortex core. \Fref{fig:figure7a_7f} illustrates this phenomenon for various microswimmer shapes. The saddle point plays an important role in this context: the maximum expansion direction at this fixed point leads to enhanced escape rates, allowing microswimmers to escape the vortex core faster. For a constant $\veff$, we observe that oblate ellipsoids escape the homoclinic orbit faster than prolate ellipsoids. This is due to the fact that for constant $\veff$ prolate microswimmers swim slower than oblate microswimmers.

\section{Summary and conclusions}

We have studied self-propelled ellipsoidal particles as idealized microswimmers in a two-dimensional axisymmetric vortex flow. In particular, we have investigated under which conditions microswimmers are trapped by the vortex, and whether they exhibit preferential orientation. This simple setting reveals interesting insights:
due to the axisymmetry of the problem, the phase space is two-dimensional and can be parameterized by the microswimmer's radial position and an orientation angle (relative to the position vector). Topologically, the phase space features a saddle point and a center. Microswimmers bound to the vortex core follow closed orbits inside a homoclinic orbit. Clustering in the laboratory frame occurs as a consequence of phase-space contraction. Shape plays a decisive role: for spherical microswimmers, we have shown that the dynamics are Hamiltonian, excluding clustering as a result of phase-space conservation. However, non-spherical particles break the Hamiltonian structure, hence enabling phase-space contraction and shape-induced clustering.
\newline
To determine whether microswimmers are trapped, we identified the effective swimming velocity as the central control parameter. The effective swimming velocity depends both on the swimming velocity and on the shape: at a constant swimming velocity, prolate ellipsoids have a larger effective swimming velocity than oblate ellipsoids. This allowed us to map the dynamics of microswimmers with different shapes to topologically equivalent phase spaces.
Using bifurcation analysis, we determined the maximum velocity for a given flow profile such that microswimmers faster than this velocity are fast enough to escape the vortex core. Notably, this maximum velocity is lower than the maximum azimuthal flow velocity, implying that microswimmers with a smaller swimming velocity than the advecting flow field can escape the vortex. Shape plays a role also here as prolate microswimmers can more easily escape than oblate microswimmers.
As the shapes of many plankton and bacteria species can be approximated as thin rods \cite{kaya2009prl}, this effect may have implications for motile species in aquatic environments. In particular, a prolate shape may yield advantages such as avoiding hydrodynamic trapping while grazing or escaping predation, without having to dedicate additional energy to swim faster.\newline
Finally, we investigated the impact of rotational noise. We find that the inclusion of rotational noise allows for initially trapped microswimmers to escape the vortex core. The presence of the saddle point leads to enhanced escape rates as the maximum expansion direction quickly drives microswimmers away from the vortex core.\newline
While we focused on the simple case of an axisymmetric flow, our results might help to also better understand the impact of shape and motility on microswimmer dynamics in more complex flows.

\ack
This work was supported by the Max Planck Society. MW also gratefully acknowledges a Fulbright-Cottrell Award grant. The authors thank Ramin Golestanian for helpful discussions.
\\
\section*{References}
\bibliography{main}
\clearpage

\end{document}